\documentclass{iaus}
\usepackage{graphicx}

\title[The ESSENCE Supernova Survey]      
{Type Ia Supernovae and the Acceleration \\ of the Universe: Results from
the \\ ESSENCE Supernova Survey}

\author[Kevin Krisciunas]   
{Kevin Krisciunas$^1$}

\affiliation{$^1$Department of Physics, Texas A\&M University, \\ 4242 TAMU,
College Station, Texas 77843, USA\\ email: {\tt krisciunas@physics.tamu.edu}}

\pubyear{2008}
\jname{MEARIM $-$ 1$^{st}$ Meeting}
\editors{A. Hady, ed.}

\begin{document}

\maketitle

\begin{abstract}
The ESSENCE project was a six year supernova search carried out with the
CTIO $\;$ 4-m telescope.  We also obtained spectra with many of the world's
largest ground-based telescopes and observed some of our SNe with the
Hubble Space Telescope and the Spitzer Space Telescope.  We achieved our
goal of discovering over 200 Type Ia SNe in the redshift range 0.2 to 0.8.
With these data we determined the cosmic equation of state parameter to 
$\pm$ 10 percent.  The data are consistent with a geometrically flat universe
whose dark energy is equivalent to Einstein's cosmological constant.
 
\end{abstract}

\firstsection 
\section{Cosmology with Type Ia Supernovae}

In the early- to mid-1990's astronomers expected that observations of standardizable
candles such as Type Ia supernovae (SNe) would reveal to what extent the expansion of
the universe was being decelerated by the gravitational attraction of all the matter in
it.  Some astronomers predicted that $\Omega _M \equiv \rho _0 / \rho _{crit}$ = 1, the
so-called Einstein-de Sitter universe.  From the dynamics of clusters of galaxies and
studies of the large scale structure of the universe, others believed that $\Omega _M
\approx$ 0.2-0.3.  Then \cite[Riess \etal\ (1998)]{Riess_etal98} and \cite[Perlmutter
\etal\ (1999)]{Perlmutter_etal99} independently showed that Type Ia SNe at a redshift
of $z \sim$ 0.5 were systematically ``too faint'' and that their faintness could be
attributed to an acceleration of the universe caused by a non-zero vacuum energy
density.  We refer to this as dark energy.

\cite[Einstein (1917)]{Einstein17} introduced the cosmological constant ($\Lambda$) to
characterize a universe that is neither expanding nor contracting.  (This was 12 years
before Hubble's discovery of the expansion of the universe.)  Modern cosmologists
describe the vacuum energy density by the dimensionless parameter $\Omega _{\Lambda}
\equiv \Lambda c^2 / 3 (H_0)^2$, where H$_0$ is Hubble's constant. 
1/$\sqrt \Lambda$ has units of length, the ``length scale over which the gravitational
effects of a nonzero vacuum energy density would have an obvious and highly visible
effect on the geometry of space and time'' (\cite[Abbott 1988]{Abbott88}).  For H$_0$ =
72 km sec$^{-1}$ Mpc$^{-1}$ and $\Omega _{\Lambda}$ = 0.7, the length scale is 2900
Mpc, more than half the size of the observable universe.

Supernova data, combined with the observations of the baryon acoustic oscillations
(\cite[Eisenstein \etal\ 2005]{Eisenstein_etal05}) or combined with WMAP satellite
observations of the cosmic microwave background radiation (\cite[Komatsu \etal\
2008]{Komatsu_etal08}), indicate that $\Omega _M + \Omega _{\Lambda}$ = 1.  In other
words, the geometry of the universe is flat.  The expansion of the universe was
slowing down for the first $\sim$7 billion years after the Big Bang.  Since then it has
been accelerating.

In order to investigate the nature of dark energy, we wish to characterize its equation
of state.  Let the equation of state parameter $w \equiv P/(\rho c^2)$, where $P$ is the pressure
and $\rho$ is the energy density.  $w$ = +1/3 for radiation.  $w = -1$ for Einstein's
cosmological constant.  However, more exotic possibilities are possible.  $w > -1$
could indicate cosmic strings, and $w < -1$ could lead to the eventual instability of
all particles in the universe (the so-called ``Big Rip'').

The luminosity distance, measured in Mpc, is given by

\begin{displaymath}
d_{lum} = \frac{c(1+z)}{H_o}\int_{0}^{z}\frac{dz'}
    {\sqrt{ \Omega _M(1 + z')^3 + \Omega _{\Lambda}(1 + z')^{3(1+w)} } } \;\; .
\end{displaymath}

The distance modulus $\mu \equiv m-M$ = 5 log ($d_{lum}$) + 25.  A plot of the distance
moduli of supernovae vs. the redshifts (or logarithms of the redshifts) is called the
Hubble diagram.  Beyond a redshift of $z \sim$ 0.2 the loci characterized by various
combinations of $\Omega _M$, $\Omega _{\Lambda}$, and $w$ begin to diverge from each
other.  If we can obtain accurate distance moduli and spectroscopic redshifts of SNe or
their host galaxies, we should in principle be able to determine these parameters.  It
is also customary to pick one locus in the Hubble diagram as a reference and to plot
the differences of the distance moduli with respect to that locus.

Determining the distance modulus of a given SN involves several steps: 1) imaging it in
multiple filters; 2) transforming the photometry to observer rest frame apparent magnitudes; 3)
determining the apparent magnitude(s) at maximum light;  4) correcting the apparent
magnitudes for dust extinction along the line of sight; and 5) subtracting an estimate
of the supernova's absolute magnitude at maximum for each filter.  Since the discovery
of the brightness/decline rate relation by \cite[Phillips (1993)]{Phillips93} and the
Cepheid-based calibration, using the Hubble Space Telescope, of the distances of nearby
galaxies which have hosted Type Ia SNe (\cite[Suntzeff \etal\ 1999]{Suntzeff_etal99},
\cite[Freedman \etal\ 2001]{Freedman_etal01}),
we have come to a greater understanding of the uniformity of the light curves and
absolute magnitudes of these objects.  It is important to note that without
observations in at least two filters, no host galaxy extinction corrections can be
made.  Supernovae observed in only one filter should not be used for the determination
of cosmological parameters. Observations of a Type Ia SN in four or more rest frame
filters, especially if one is a near-infrared filter, allow an accurate determination
of the dust extinction suffered by the SN, even if the dust is very unusual
(\cite[Krisciunas \etal\ 2007]{Krisciunas_etal07}).

\section{The ESSENCE Supernova Survey}

In order to measure the equation of state parameter to $\pm$ 10 percent we organized the 
ESSENCE supernova survey.  \cite[Miknaitis \etal\ (2007)]{Miknaitis_etal07} describe
the motivation and logistics of the survey.

ESSENCE stands for {\bf E}quation of {\bf S}tate:  {\bf S}up{\bf E}er{\bf N}ovae trace
{\bf C}osmic {\bf E}xpansion.  The ESSENCE team presently consists of 32 astronomers
from the United States, Chile, Germany, Sweden, and Australia.  We observed for six
seasons (October through December) with the 4-m telescope of the Cerro Tololo
Inter-American Observatory, using the prime focus Mosaic II camera. The camera contains
eight 2K $\times$ 4K CCD chips, each read out with two amplifiers. The field of view is
36 $\times$ 36 arcmin.  We had 32 principal fields. Observing every other half-night
during dark and grey time, we typically observed 16 fields each night.

Over the course of the survey we observed on 191 nights and took 5458 $R$- and
$I$-band images.  Depending on the redshift of the SN, this corresponded to rest frame
$UB$ or rest frame $BV$ imaging.  We found over 2000 flux transients.  Using the
Magellan 6.5-m telescopes, the Gemini 8-m telescopes, the Very Large Telescope, and
Keck, we took spectra of 400 targets.  Some of these objects turned out to be Active
Galactic Nuclei, supernovae other than Type Ia, or variable stars in our Galaxy.  
Roughly 220 targets were confirmed to be Type Ia SNe or possible Type Ia SNe.

Analysis of the spectra is ongoing. \cite[Matheson \etal\ (2005)]{Matheson_etal05}
discussed spectra of the first two years of the survey.  \cite[Blondin \etal\ 
(2006)]{Blondin_etal06} used line profile shapes to test the fraternity of Type Ia SNe
at high and low redshifts.  \cite[Foley \etal\ (2008)]{Foley_etal08} showed that
there is no strong evidence for evolution of these objects out to redshift $z$ = 0.8.
\cite[Bronder \etal\ (2008)]{Bronder_etal08} and \cite[Ellis \etal\ (2008)]{Ellis_etal08}
independently came to the same conclusion using spectra from the Supernova Legacy
Survey (SNLS), a supernova search being carried out year-round on the 3.6-m 
Canada-France-Hawaii Telescope at Mauna Kea.

Some of our SNe were also observed with HST or with the Spitzer Space Telescope.  
\cite[Krisciunas \etal\ (2005)]{Krisciunas_etal05} discuss nine ESSENCE SNe observed
with HST in three rest frame optical bands.  Most of these turned out to be very slowly
declining (i.e. overluminous) objects.  In the 2006 and 2007 observing seasons we used
HST to observe ten ESSENCE SNe with $z \approx$ 0.35 at 1.65 $\mu$m, which is
to say in the rest frame $J$-band (1.25 $\mu$m).

\begin{figure}[b]
\begin{center}
 \includegraphics[width=3.4in]{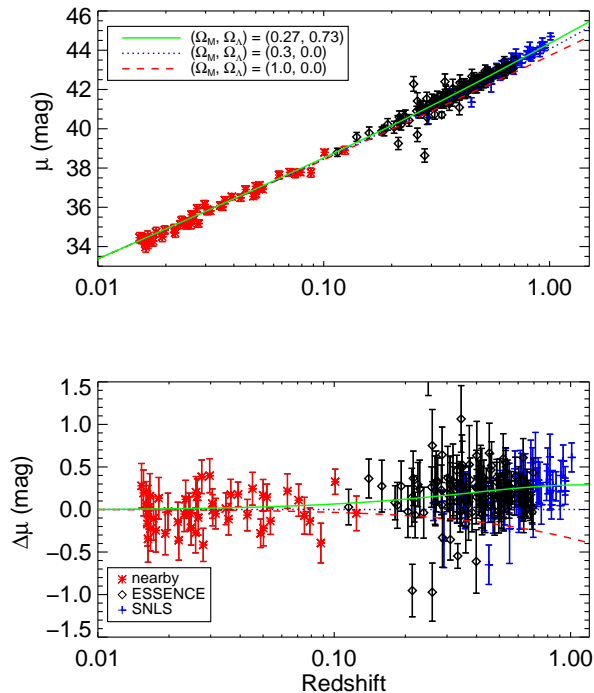} 
 \caption{{\em Top}: Preliminary Hubble diagram of $\sim$200 ESSENCE Type Ia SNe (black
diamonds), nearby objects (red asterisks), and SNe from the first year of
the Supernova Legacy Survey (blue pluses).  The solid green line, with $\Omega _M$ =
0.27, $\Omega _{\Lambda}$ = 0.73, gives the best fit.  Also shown are two models
with zero cosmological constant ($\Omega _M$ = 0.3 and 1.0, respectively).
{\em Bottom}: Differential Hubble diagram using the ``open'' model ($\Omega _M$ = 0.3, 
$\Omega _{\Lambda}$ = 0.0) as reference.}
   \label{fig1}
\end{center}
\end{figure}

Using data from the first three years of the ESSENCE survey \cite[Wood-Vasey \etal\
(2007)]{Wood-Vasey_etal07} found that $w = -1.07 \pm 0.09$ (statistical) $\pm$ 0.13
(systematic).  We found $\Omega _M = 0.267^{+0.028}_{-0.018}$.  The first year SNLS
data (\cite[Astier \etal\ 2006]{Astier_etal06}) lead to very similar values:  $w =
-1.023 \pm 0.090$ (statistical) $\pm$ 0.054 (systematic); $\Omega _M = 0.263 \pm
0.042$.

In Fig. \ref{fig1} we show the preliminary Hubble diagram and a differential Hubble
diagram of our $\sim$200 ESSENCE Type Ia SNe, along with nearby objects and those from 
the first year of SNLS.  We clearly have some outliers.  These could be due to ``bad''
photometry, incorrect extinction corrections, or incorrect redshifts.  Some of these
objects could be members of new sub-classes of SNe, so the adopted absolute magnitudes
are wrong. Still, the data are consistent with a flat universe containing dark energy.  
Our final value of the equation of state parameter requires further analysis, but will
likely be statistically consistent with $w = -1$.

\section{Other Results from the Collaboration}

The ESSENCE database has enabled other interesting research.
\cite[Davis \etal\ (2007)]{Davis_etal07} scrutinized exotic cosmological models using
ESSENCE SNe and higher-$z$ objects discovered by \cite[Riess \etal\
(2004)]{Riess_etal04}.  We found that, ``the preferred cosmological model is the flat
cosmological constant model, where the expansion history of the universe can be
adequately described with only one free parameter describing the energy content of the
universe. Among the more exotic models that provide good fits to the data, we
note a preference for models whose best-fit parameters reduce them to the cosmological
constant model.''

\cite[Blondin \etal\ (2008)]{Blondin_etal08} investigated time dilation effects in
multi-epoch high redshift Type Ia SN spectra. By comparing the rest frame age of
each spectrum (determined through cross-correlations with a database of
spectral templates) with the observed elapsed time, we found an aging rate
consistent with the expected $1/(1+z)$ factor.

ESSENCE discovered {\em moving} objects in our solar system as well. \cite[Becker
\etal\ (2008)]{Becker_etal08} report the discovery of 15 trans-Neptunian objects.
Several have orbits which are highly elliptical (up to 0.85) or substantially inclined
to the plane of the ecliptic.  Two have aphelia of 352 and 582 AU, respectively.


\begin{thebibliography}{}

\bibitem[Abbott (1988)]{Abbott88}
{Abbott, L., } 1988, {\em Sci. Amer.}, 258, no. 5, 106

\bibitem[Astier \etal\ (2006)]{Astier_etal06}
{Astier, P., et al.} 2006, \textit{A\&A}, 447, 31

\bibitem[Becker \etal\ (2008)]{Becker_etal08}
{Becker, A. C., et al.} 2008, \textit{ApJ}, 682, L53

\bibitem[Blondin \etal\ (2006)]{Blondin_etal06}
{Blondin, S., et al.} 2006, \textit{AJ}, 131, 1648

\bibitem[Blondin \etal\ (2008)]{Blondin_etal08}
{Blondin, S., et al.} 2008, \textit{ApJ}, 682, 724

\bibitem[Bronder \etal\ (2008)]{Bronder_etal08}
{Bronder, T. J., et al.} 2008, \textit{A\&A}, 477, 717

\bibitem[Davis \etal\ (2007)]{Davis_etal07}
{Davis, T., et al.} 2007, \textit{ApJ}, 666, 716

\bibitem[Einstein (1917)]{Einstein17}
{Einstein, A.,} 1917, Sitzungsberichte der K\"{o}niglich Preuss. Akad. der Wissenschaften, 142

\bibitem[Eisenstein \etal\ (2005)]{Eisenstein_etal05}
{Eisenstein, D., et al.} 2005, \textit{ApJ}, 633, 560

\bibitem[Ellis \etal\ (2008)]{Ellis_etal08}
{Ellis, R. S., et al.} 2008, \textit{ApJ}, 674, 51

\bibitem[Foley \etal\ (2008)]{Foley_etal08}
{Foley, R., et al.} 2008, \textit{ApJ}, in press, 684, 68

\bibitem[Freedman \etal\ (2001)]{Freedman_etal01}
{Freedman, W., et al.} 2001, \textit{ApJ}, 553, 47

\bibitem[Komatsu \etal\ (2008)]{Komatsu_etal08}
{Komatsu, E., et al.} 2008, \textit{ApJ}, in press, arXiv:0803.0547

\bibitem[Krisciunas \etal\ (2005)]{Krisciunas_etal05}
{Krisciunas, K., et al.} 2005, \textit{AJ}, 130, 2453

\bibitem[Krisciunas \etal\ (2007)]{Krisciunas_etal07}
{Krisciunas, K., et al.} 2007, \textit{AJ}, 133, 58

\bibitem[Matheson \etal\ (2005)]{Matheson_etal05}
{Matheson, T., et al.} 2005, \textit{AJ}, 129, 2352

\bibitem[Miknaitis \etal\ (2007)]{Miknaitis_etal07}
{Miknaitis, G., et al.} 2007, \textit{ApJ}, 666, 674

\bibitem[Perlmutter \etal\ (1999)]{Perlmutter_etal99}
{Perlmutter, S., et al.} 1999, \textit{ApJ}, 517, 565

\bibitem[Phillips (1993)]{Phillips93}
{Phillips, M. M.,} 1993, \textit{ApJ}, 413, L105

\bibitem[Riess \etal\ (1998)]{Riess_etal98}
{Riess, A. G., et al.} 1998, \textit{AJ}, 116, 1009

\bibitem[Riess \etal\ (2004)]{Riess_etal04}
{Riess, A. G., et al.} 2004, \textit{AJ}, 607, 665

\bibitem[Suntzeff \etal\ (1999)]{Suntzeff_etal99}
{Suntzeff, N. B., et al.} 1999, \textit{AJ}, 117, 1175

\bibitem[Wood-Vasey \etal\ (2007)]{Wood-Vasey_etal07}
{Wood-Vasey, W. M., et al.} 2007, \textit{ApJ}, 666, 694



\end{thebibliography}
\end{document}